\documentclass[12pt]{article}

\usepackage{amsmath}
\usepackage{amsfonts}
\usepackage{amssymb}
\usepackage{graphicx}
\usepackage{psfrag}
\usepackage{graphicx}
\usepackage{psfrag}

\newcommand{\be}{\begin{equation}}
\newcommand{\ee}{\end{equation}}
\newcommand{\ba}{\begin{eqnarray}}
\newcommand{\ea}{\end{eqnarray}}

\begin{document}
\begin{center}
{\bf
ZERO ENERGY STATES FOR A CLASS OF TWO-DIMENSIONAL POTENTIALS IN GRAPHENE
}\\
\vspace{1cm}
{\large M. V. Iof\/fe$^{1,}$\footnote{E-mail: m.ioffe@spbu.ru; corresponding author},
D. N. Nishnianidze}$^{2,}$\footnote{E-mail: cutaisi@yahoo.com}\\
\vspace{0.5cm}
$^1$ Saint Petersburg State University, 7/9 Universitetskaya nab., St.Petersburg, 199034 Russia.\\
$^2$ Akaki Tsereteli State University, 4600 Kutaisi, Georgia.\\

\end{center}
\vspace{0.5cm}
\hspace*{0.5in}
\vspace{1cm}
\hspace*{0.5in}
\begin{minipage}{5.0in}
{\small
The excitations in graphene and some other materials are described by two-dimensional massless Dirac equation with applied external potential of some kind. Solutions of this zero energy equation are built analytically for a wide class of scalar potentials. In contrast to most publications on analytical solutions of massless two-dimensional Dirac equation, our  potentials really depend on both spatial coordinates in some bounded domain. Several examples of such construction are given explicitly.
}
\end{minipage}



%\section*{\normalsize\bf 1. \quad Introduction.}
\section{Introduction.}
%\vspace*{0.5cm}
%\hspace*{3ex}
During last years, different properties of graphene \cite{novoselov}-\cite{kats} - the atomically thin conducting material with carbon atoms in a honeycomb lattice - were investigated extensively by many authors. In particular, the behaviour of electron carriers within graphene in the presence of external fields was studied for the external fields of different nature and configuration \cite{efetov-1}, \cite{efetov-2}, \cite{matulis}, \cite{jakubsky-2}, \cite{jakubsky-3}.
%\cite{jakubsky-1}, \cite{fernandez-1}.
It is known that in the tight-binding approximation, the excitations near the Fermi surface correspond to solutions of two-dimensional Dirac equation with zero mass (i.e. with zero energy) \cite{novoselov}, \cite{novoselov-2}. Thus, the special attention was attracted to solutions of this relativistic equation with different forms of interaction terms \cite{peres},
%\cite{kuru},
\cite{bardarson}, \cite{portnoi-4},
%\cite{milpas},
\cite{portnoi-1}, \cite{fernandez-2}, \cite{portnoi-2}, \cite{ho-1}, \cite{ghosh}, \cite{ho-2},   \cite{portnoi-3}, \cite{portnoi-11},\cite{portnoi-111}, \cite{schulze} (see also on two-particle zero-energy states in \cite{2-part}. In this context, the up and down elements of the Dirac two-component spinor are the wave functions corresponding to two different sublattices in graphene.

In the variety of papers on two-dimensional Dirac equation with scalar external potential, the latter was usually chosen depending only on one spatial variable $V(x_1).$ This essential restriction allowed to replace the initial Dirac equation, i.e. a pair of first order differential equations, by the corresponding second order one-dimensional equation of the Schr\"odinger form. Due to special choice of exactly solvable potential $V(x_1),$ this $x_1-$ dependent equation can be solved analytically leading to the normalizable part of wave function $\Psi (x_1).$ In its turn, the $x_2-$dependent equation is trivial - it corresponds to the free motion with plane wave eigenfunction $\Psi(x_2)=\exp(ikx_2).$ In the present paper, we get rid of this requirement of the one-dimensionality of the potential. Instead, we study potentials which depend on both coordinates $x_1,\, x_2,$ and corresponding two-dimensional Dirac equation will be solved explicitly for a wide class of such potentials. The paper is organized as follows. After formulation of the problem in terms of second order differential equation for the component $\Psi_A(\vec x)$ of "spinor" from two-dimensional Dirac equation (Section 2), it is solved analytically for specific particular case when the first derivative can be removed from this Schr\"odinger-like equation (Section 3). This condition provides the opportunity to replace the physical coordinates $(x_1,x_2)$ by new variables $(\tau_1,\tau_2)$ which allow the procedure of separation.  As a result, the massless Dirac equation is exactly solved in terms of a  holomorphic function $f(z)$ which defines the two-dimensional potential $V(x_1,x_2).$ Section 4 presents several examples of proposed algorithm, and some conclusions are given in Section 5.

\section{Formulation of the problem.}

The considered problem is the following. We start with mentioned above zero energy two-dimensional Dirac equation:
\be
(\sigma_1p_1 + \sigma_2p_2 + V(\vec x))\Psi(\vec x)=0,   \label{dirac}
\ee
where the Fermi velocity was taken unity, $\sigma_1, \sigma_2$ are standard Pauli matrices, $x_1, x_2$ - spatial coordinates, $p_1, p_2$ - corresponding momenta, $V(\vec x)$ - real potential, and $\Psi(\vec x)$ is a two-component column - "spinor" - with components $\Psi_A(\vec x), \Psi_B(\vec x).$ It is convenient to introduce mutually conjugate complex variables $z, \bar z$ and corresponding derivatives $\partial\equiv \frac{\partial}{\partial z}, \bar\partial \equiv \frac{\partial}{\partial \bar z} :$
\begin{equation}\label{z}
z\equiv x_1+ix_2;\quad \bar z\equiv x_1-ix_2;\quad \partial \equiv \frac{1}{2}(\partial_1-i\partial_2);\quad \bar\partial \equiv \frac{1}{2}(\partial_1+i\partial_2),
\end{equation}
so that Eq.(\ref{dirac}) can be rewritten as a system of coupled equations:
\begin{eqnarray}
% \nonumber to remove numbering (before each equation)
  V\Psi_A - 2i\partial \Psi_B &=& 0, \label{A}\\
  V\Psi_B-2i\bar\partial\Psi_A &=& 0. \label{B}
\end{eqnarray}
The unknown function $\Psi_B$ can be eliminated:
\begin{equation}\label{psi-B}
  \Psi_B = 2iV^{-1}\bar\partial\Psi_A,
\end{equation}
and we have the second order equation for $\Psi_A:$
\begin{equation}\label{psi-A}
[4V\bar\partial\partial-4(\partial V)\bar\partial +V^3]\Psi_A=0.
\end{equation}
This equation together with Eq.(\ref{psi-B}) is equivalent to the initial Dirac system (\ref{dirac}).

\section{Solution of Dirac equation.}

Now, we restrict the model by condition that the first derivative can be excluded from the operator in (\ref{psi-A}). For this, we represent $\Psi_A$ as a product:
\begin{equation}\label{product}
\Psi_A(z,\bar z)\equiv g(z,\bar z)\varphi(z,\bar z)
\end{equation}
with function $g(z,\bar z)$ providing an absence of the first derivatives in equation for $\varphi :$
\begin{eqnarray}
\bar\partial g(z, \bar z) &=& 0, \label{gg} \\
 V\partial g-g\partial V &=& 0, \label{Vg}
\end{eqnarray}
i.e. $g$ is a holomorphic function $g=g(z),$ and $\partial(g(z)/V(z,\bar z))=0.$ Denoting the latter fraction as $g(z)/V(z, \bar z)\equiv \gamma (\bar z),$ from the reality of potential $V=\overline{V},$ one obtains the following:
\begin{equation}\label{reality}
g(z)\bar\gamma(z)=\bar g(\bar z)\gamma(\bar z)\equiv c^{-1};\quad V(z,\bar z)=c |g(z)|^2,
\end{equation}
where $c$ is an arbitrary real number. Eq.(\ref{psi-A}) gives the equation for the function $\varphi ,$ which has the form of zero energy Schr\"odinger equation with potential $-V^2=-c^2\mid g(z)\mid^4:$
\begin{equation}\label{varphi}
[4\partial\bar\partial + V^2(z,\bar z)]\varphi(z,\bar z)=0.
\end{equation}
This equation can be solved explicitly by transition to new variables:
\begin{equation}\label{tau}
\tau_1\equiv \int g^2(z)dz + \int \bar g^2(\bar z)d\bar z;\quad \tau_2\equiv i\int \bar g^2(\bar z)d\bar z - i\int g^2(z)dz,
\end{equation}
which are correspondingly the real and imaginary parts of an arbitrary function:
\begin{equation}\label{f}
f(z)\equiv \tau_1+i\tau_2 = 2\int g^2(z)dz.
\end{equation}
Indeed, in new variables, Eq.(\ref{varphi}) takes the form of two-dimensional Schr\"odinger equation for free particle with an eigenvalue $c^2/4$:
\begin{equation}\label{main}
(-\partial_{\tau_1}^2 - \partial_{\tau_2}^2)\varphi(z,\bar z) = \frac{1}{4}c^2 \varphi(z,\bar z).
\end{equation}
One has to remember that the physical coordinate space is still $(x_1, x_2)$, the first component of wave function is $\Psi_A(z,\bar z)=g(z)\varphi(z,\bar z),$ and the second component $\Psi_B$ is defined by (\ref{psi-B}).

The general solution of Eq.(\ref{main}) is obtained as an arbitrary linear combination of functions:
\begin{equation}\label{solution}
\varphi_{\vec k}(z,\bar z)=\varphi_{k_1}^{(1)}(\tau_1)\varphi_{k_2}^{(2)}(\tau_2);\quad \vec k\equiv (k_1,k_2),
\end{equation}
where
\begin{equation}\label{partial}
\varphi_{k_i}(\tau_i)\equiv \sigma_{k_i}\exp(-k_i\tau_i)+\delta_{k_i}\exp(+k_i\tau_i); \quad i=1,2,
\end{equation}
$\sigma_{k_i},\, \delta_{k_i}$ are constants, and constants $k_i,$ in general - complex, must ensure $k_1^2+k_2^2= - c^2/4$ with real $c.$

The norm of wave function $\|\Psi(x_1, x_2)\|^2=\|\Psi_A(x_1, x_2)\|^2 + \|\Psi_B(x_1, x_2)\|^2$ consists of two terms where:
\ba
\|\Psi_A\|^2&=&\int |\Psi_A|^2dx_1dx_2=\int |\varphi_{k_1}(\tau_1)|^2 |\varphi_{k_2}(\tau_2)|^2 |g(z)|^2 |\frac{\partial(x_1,x_2)}{\partial(\tau_1,\tau_2)}| d\tau_1 d \tau_2= \nonumber\\
&=&\int |\varphi_{k_1}(\tau_1)|^2 |\varphi_{k_2}(\tau_2)|^2 (f^{\prime}(z)\bar f^{\prime}(\bar z))^{-1/2} d\tau_1 d \tau_2, \label{norm}
\ea
and the corresponding expression for $\|\Psi_B(x_1, x_2)\|^2$ with (see Eq.(\ref{psi-B}))
\begin{equation}\label{ppsi-B}
\Psi_B(z, \bar z)=\frac{2i}{c\bar g(\bar z)}\bar\partial\varphi (z,\bar z).
\end{equation}

According to the well known Liouville theorem \cite{complex}, a nontrivial holomorphic function $|f(z)|\not\equiv Const$ can not be bounded at the whole complex plane $(z,\bar z)$. Therefore, in the case of nonvanishing real part of $k_i$ in (\ref{partial}), the unbounded variables $\tau_1$ and $\tau_2$ give the function $\varphi,$ in its general form (\ref{solution}), exponentially increasing in some directions in the plane $(\tau_1, \tau_2).$ This behaviour prevents the construction of wave functions normalizable on the whole plane $(x_1,x_2).$ One special opportunity has to be mentioned: if both constants $k_i$ are pure imaginary, the norm (\ref{norm}) reduces
to the $|\int dz g(z)|^2.$ Again due to Liouville theorem, on the whole plane this expression is unbounded. It is appropriate to recall here that all solutions of Eq.(\ref{dirac}) built for different models of {\it one-dimensional} potentials $V(x_1)$ also obey the analogous property: they include the plane wave in $x_2-$direction \cite{peres},
%\cite{kuru},
\cite{portnoi-4},
%\cite{milpas},
\cite{fernandez-2}, \cite{portnoi-2}, \cite{ho-1}, \cite{ghosh}, \cite{ho-2}, \cite{schulze}. In our case of {\it two-dimensional} potentials, one can however use the solutions obtained in the previous Section for some domain of the plane in variables $(\tau_1, \tau_2)$ with suitable boundary conditions. The condition for the choice is that in this domain each variable $\tau$ is either finite or takes only one of infinite values ($\pm\infty$). In this case, functions $\varphi$ in (\ref{partial}) have finite $|\varphi |$, and therefore, lead to normalizable wave functions $\Psi$ due to (\ref{norm}) even if constants $k_i$ have nonvanishing real part. Correspondingly, each such region in the plane $(\tau_1,\tau_2)$ has its own prototype in the physical coordinate space $(x_1,x_2).$ Several examples of this construction are given below.

\section{Examples.}

1.\quad For the first example, we choose an arbitrary power function $f(z)=z^n$ with $n>1.$ It is convenient here to consider the physical plane $(x_1,x_2)$ in polar coordinates $(\rho, \theta).$ Then, the variables $\tau_1, \tau_2$ are:
\begin{equation}\label{power}
\tau_1=\rho^n\cos(n\theta);\quad \tau_2=\rho^n\sin(n\theta).
\end{equation}
It is clear that in the domain bounded by two rays $\theta \in [0, \frac{\pi}{2n}]$ and arbitrary radius $\rho,$ the variables $\tau$ are nonnegative and belong to the first quadrant:
\begin{equation}\label{rays}
\tau_1 \geq 0; \quad \tau_2 \geq 0.
\end{equation}
This is just the case when solutions $\varphi$ of (\ref{partial}) with nonvanishing real part of $k_i$ can be chosen such that they do not diverge anywhere in this quadrant, i.e. in the physical terms, in the sector restricted by $\theta \in [0, \frac{\pi}{2n}].$ The vanishing at the origin multiplier $(f^{\prime}(z)\bar f^{\prime}(\bar z))^{-1/2}$ in the integrand in (\ref{norm}) is compensated by the measure. According to (\ref{reality}), the potential for this example has the form:
\begin{equation}\label{V-1}
V=\frac{c}{2}|f^{\prime}(z)|=\frac{cn}{2}\rho^{n-1},
\end{equation}
and the boundary values of solutions are defined by the coefficients $\sigma_{k_i}$ in (\ref{partial}). In particular, choosing only one nonzero pair of mutually conjugated values $k_1=k_2^{\star}$ and all $\delta_{k_i}\equiv 0,$ one obtains explicitly the wave function
$$\Psi_A(\vec x)=g(z) \varphi_{\vec k}(\vec \tau )=(f'(z)/2)^{1/2}\varphi_{k_1}(\tau_1)\varphi_{k_1^{\star}}(\tau_2) =\sqrt{n/2}\sigma_{k_1}\sigma_{k_1^{\star}}z^{(n-1)/2}\exp{(-k_1\tau_1-k_1^{\star}\tau_2)}.$$ This wave function in the sector $\theta \in [0, \frac{\pi}{2n}]$ of the physical plane $(x_1, x_2)$ satisfies the boundary conditions quasi-periodic in angle $\theta ,$ i.e. $\Psi_A(\rho, \theta =\pi/2n)=\Psi_A(\rho, \theta =0),$ but up to a phase factor.
In the case of pure imaginary $k_1=k_2,$ after choosing again $\delta_{k_i}\equiv 0,$ the same boundary problem is solved even in a simpler way.
One can also try the functions $f(z)$ of more general polynomial forms which lead to potentials non-invariant under rotations.

%\vspace{6pt}

%\begin{center}
%\includegraphics[height=5cm]{1.pdf}%{rc.eps}

%\noindent{\it Fig.1} Plot
%of the potential $V^{(1)}$ (\ref{44}) for $ k_1=k_2=0,\, k=1,\, \lambda =2.$

%\end{center}

%\vspace{4pt}

%2.\quad For the second example, we take the polynomial function:
%\begin{equation}\label{a}
%f(z)=az^3+bz
%\end{equation}
%with positive constants $a, b.$ Then, the variables $\tau_1, \tau_2$ are:
%\begin{equation}\label{power+}
%\tau_1=ax_1(x_1^2-3x_2^2)+bx_1;\quad \tau_2=ax_2(3x_1^2-x_2^2)+bx_2.
%\end{equation}
%One can check that the domain bounded by two rays $\theta \in [0, \frac{\pi}{6}]$ and arbitrary radius $\rho,$ the variables $\tau$ are nonnegative and again belong to the first %quadrant (\ref{rays}) ????:
%\begin{equation}\label{rays+}
%\tau_1 \geq 0; \quad \tau_2 \geq 0.
%\end{equation}
%In polar coordinates, the corresponding potential is not rotationally symmetric now:
%\begin{equation}\label{V-1+}
%V=c|f^{\prime}(z)|^2=\frac{c}{4}(9a^2\rho^4+6ab\rho^2\cos(2\theta)).
%\end{equation}

2.\quad Let us take:
\begin{equation}\label{th}
f(z)=a \tanh(\lambda z)
\end{equation}
with positive constants $a, \lambda.$ In this case,
\begin{equation}\label{ff}
\tau_1=a\frac{\sinh(2\lambda x_1)}{\cosh(2\lambda x_1) + \cos(2\lambda x_2)};\quad
\tau_2=a\frac{\sin(2\lambda x_2)}{\cosh(2\lambda x_1) + \cos(2\lambda x_2)}.
\end{equation}
These relations provide that the first quadrant in the plane $(\tau_1, \tau_2)$ corresponds now to a half-strip in the physical variables $(x_1, x_2).$ The border of the region is described as $x_1\in [0, \infty ); \quad x_2 \in [0, \frac{\pi}{2\lambda}].$ In this example, the potential has the form:
\begin{equation}\label{V-2}
V(z,\bar z)= \frac{c a\lambda }{2}|\cosh(\lambda z)|^{-2}=\frac{c a\lambda}{\cosh(2\lambda x_1)+\cos(2\lambda x_2)}.
\end{equation}

%4.\quad The function
%\begin{equation}\label{tth}
%f(z)=a \cosh(\lambda z)
%\end{equation}
%with real constants $a, \lambda$ in the domain:
%\begin{equation}\label{ffff}
%\tau_1=a\cosh(\lambda x_1)\cos(\lambda x_2)\geq 0;\quad
%\tau_2=a\sinh(\lambda x_1)\sin(\lambda x_2)\geq 0
%\end{equation}
%also corresponds to the same half-strip $x_1\in [0, \infty ); \quad x_2 \in [0, \frac{\pi}{2\lambda}],$ but with another potential:
%\begin{equation}\label{V-3}
%V(z,\bar z)= \frac{ca^2\lambda^2}{4}|\sinh(\lambda z)|^2=\frac{ca^2\lambda^2}{8}(\cosh(2\lambda x_1)-\cos(2\lambda x_2)).
%\end{equation}
%The replacement of $\lambda $ by $ i\lambda $ in the last relations provides us with new potential in the same domain.

3. \quad Choosing $f(z)$ in the form:
\begin{equation}\label{thhh}
f(z)=a \cosh^{-1}(\lambda z)
\end{equation}
with positive constants $a, \lambda,$ we obtain:
\begin{equation}\label{fff}
\tau_1=a\frac{\cosh(\lambda x_1)\cos(\lambda x_2)}{\cosh^2(\lambda x_1) - \sin^2(\lambda x_2)};\quad
\tau_2=-a\frac{\sinh(\lambda x_1)\sin(\lambda x_2)}{\cosh^2(\lambda x_1) - \sin^2(\lambda x_2)}.
\end{equation}
In this example, the fourth quadrant in the plane $(\tau_1, \tau_2)$ corresponds to the same half-strip in the physical variables $(x_1, x_2)$ as in the previous example, but the potential has the form:
\begin{equation}\label{V-4}
V(z,\bar z)= \frac{c a\lambda }{2}|\frac{\sinh(\lambda z)}{\cosh^2(\lambda z)}|=c a\lambda \frac{(\cosh^2(\lambda x_1)-
\cos^2(\lambda x_2))^{1/2}}{\cosh(2\lambda x_1)+\cos(2\lambda x_2)}
\end{equation}

The list of such examples can be continued. For each new function $f(z),$ one obtains the new expression for potential $V(z, \bar z)$ and new form of the region in $(x_1,x_2)$ variables where the zero energy solutions of two-dimensional Dirac equation are built.

\section{Conclusions}

Summurizing, a new class of scalar potentials amenable to analytical solution of two-dimensional Dirac equation was built. In contrast to a series of previous publications, these potentials non-trivially depend on both spatial variables. Although, the specific ansatz - the absence of first derivative in Eq.(\ref{psi-A}) - was chosen, the class of such potentials is rather wide. The general solution of two-dimensional Dirac equation with these potentials was found analytically in terms of variables $\tau_1, \tau_2$ of (\ref{tau}). Similarly to the previously known (one-dimensional) potentials, where the wave functions included non-normalizable plane wave multiplier, in our case the wave functions also are not normalizable on the whole plane due to Liouville theorem and in accordance with the well known Klein paradox \cite{kats}. Therefore, a suitable region of the plane must be taken with corresponding boundary conditions. In some examples of Section 4, one quadrant in the $(\tau_1, \tau_2)-$plane was chosen as such a region.

\section{Acknowledgments}

The work of M.V.I. was partially supported by RFBR Grant No. 18-02-00264-a.

\end{document}